\newif\ifICFP
\ICFPfalse
\ifICFP
\documentclass{sig-alternate}
\else
\documentclass{article}
\fi
\usepackage[all]{xy}
\usepackage{qsymbols}
\usepackage{mathrsfs}
\usepackage{haskell}
\usepackage{enumitem}
\usepackage{fancyvrb}
\usepackage{color}
\usepackage{url}
\ifICFP\else
\usepackage{graphics}
\fi

\newqsymbol{`e}{\varepsilon}
\newcommand{\Bt}{\mathscr{B}}
\newcommand{\C}{\mathscr{C}}
\newcommand{\M}{\mathscr{M}}
\newcommand{\PP}{\mathbb{P}}
\newcommand{\Me}{\mathbb{E}}

\newcommand{\real}{\ensuremath{\mathbb{R}}}
\newcommand{\nat}{\ensuremath{\mathbb{N}}}
\newcommand{\Soo}{\ensuremath{S_{\infty}}}
\newcommand{\nbt}[1]{T_{\infty,#1}}

\newtheorem{theorem}{Theorem}

\newif\ifcomment
\commenttrue           
\definecolor{gris}{gray}{0.3}

\begin{document}
%
\ifICFP \conferenceinfo{ICFP}{International Conference on Functional Programming}

\title{Boltzmann samplers \\for random generation of lambda terms} 

\numberofauthors{1} 
%
\author{ \alignauthor
  Pierre Lescanne\\
  \affaddr{  University of Lyon}, \\
  \affaddr{\'Ecole normale sup\'erieure de Lyon,} \\
  \affaddr{LIP (UMR 5668 CNRS ENS Lyon UCBL INRIA)\\
    \affaddr{46 all\'ee d'Italie, 69364 Lyon, France}\\
    \email{pierre.lescanne@ens-lyon.fr} }}
\else
\author{Pierre Lescanne\\
   University of Lyon, \\
  {\'Ecole normale sup\'erieure de Lyon,} \\
  {LIP (UMR 5668 CNRS ENS Lyon UCBL INRIA)}\\
    {46 all\'ee d'Italie, 69364 Lyon, France}\\
    \textsf{pierre.lescanne@ens-lyon.fr} }
\title{Boltzmann samplers \\for random generation of lambda terms} 
\fi
\maketitle
\begin{abstract}
  \begin{sloppypar}
    Randomly generating structured objects is important in testing and optimizing
    functional programs, whereas generating random $`l$-terms is more specifically
    needed for testing and optimizing compilers.  For that a tool called
    \textsf{QuickCheck} has been proposed, but in this tool the
    control of the random generation is left to the programmer.  Ten years ago, a
    method called Boltzmann samplers has been proposed to generate combinatorial
    structures.  In this paper, we show how Boltzmann samplers can be developed to
    generate random \mbox{$`l$-terms}, but also other random data structures like trees.  These
    samplers rely on a critical value which parameters the main random selector and
    which is exhibited here with explanations on how it is computed.
    \textsf{Haskell} programs are proposed to show how samplers are actually
    implemented.
  \end{sloppypar}

\ifICFP\else 
 \fi
\end{abstract}

\ifICFP
\begin{sloppypar}
  \category{Theory of Computation}{Randomness, geometry and discrete
    structures}{Generating random combinatorial structures} \category{Theory of
    Computation}{Logic}{Type theory} \category{Software and its engineering}{Software
    notations and tools}{General programming languages;} { Language types;} {
    Functional languages} %
  \terms{Lambda calculus, Random Generation}
\end{sloppypar}
\fi

\section{Introduction}
Claessen and Hughes~\cite{DBLP:conf/icfp/ClaessenH00} ask a fundamental question:
\begin{it}
  \begin{quotation} [H]ow would one choose a random closed \mbox{$`l$-term} with a uniform
    distribution?
  \end{quotation}
\end{it}
Actually generating random lambda terms and more specifically generating random
typable $`l$-terms are fundamental for debugging functional programming
compilers~\cite{Palka:2011:TOC:1982595.1982615}, but also for answering many
questions concerning $`l$-calculus and functional programming.  We will see that this
is possible using an apparatus called \emph{Boltzmann samplers} (Duchon et
al.~\cite{DBLP:journals/cpc/DuchonFLS04}) and based on the ideas of statistical
physics initiated by the Austrian physicist Ludwig Boltzmann (1844-1906).  The key
issue is the ability to generate random combinatorial structures with some
flexibility on the size. In this paper the combinatorial structures are
\mbox{$`l$-terms}, but also since we are interested in functional programming they
are other data structures, like \emph{binary trees} (the data type \textsf{Tree}
in \cite{DBLP:conf/icfp/ClaessenH00}) or \emph{1-2-trees}.  These latter  structures
will allow us to introduce the concept of Boltzmann samplers.  Assume a program
generating random $`l$-terms is built.  Somewhere the program chooses whether it generates a
variable, an abstraction or an application. This is done according to a probability
distribution.  Why the chosen probability distribution is
respectively
\begin{displaymath}
  \begin{array}{l@{\qquad}l@{\qquad}l}
variable & abstraction & application\\
  0.3703026 & 0.25939476 & 0.3703026
\end{array}
\end{displaymath}
will be explained further.  For now let us say that there is a  \emph{critical value}
${`r\approx 0.509308127}$,  which plays a main role in counting
$`l$-terms and that the above values are respectively $(1-`r^2)/2$ and $`r^2$.  Once
we know how to generate large plain $`l$-terms, we are able to address the
random generation of large typable $`l$-terms.

This paper is based on four milestone papers.
\begin{enumerate}[label=\roman*]
\item \cite{NGDeBruijn108} Nicolaas Govert de~Bruijn (1972).  \emph{Lambda calculus
    notation with nameless dummies, a tool for automatic formula manipulation, with
    application to the {Church-Rosser} theorem.}  \newblock {\em Indagationes
    mathematicae}, {\bf 34}(5), 381--392.
\item \cite{DBLP:conf/icfp/ClaessenH00} Koen Claessen and John Hughes.
  \emph{\textsf{QuickCheck}: a light\-weight tool for random testing of
    \textsf{Haskell} programs.}  \newblock In Martin Odersky and Philip Wadler,
  editors, {\em ICFP}, pages 268--279. ACM, 2000. This paper received the \emph{Most
    Influential ICFP Paper Award} for the year 2000.
\item \cite{DBLP:journals/cpc/DuchonFLS04} Philippe Duchon, Philippe Flajolet, Guy
  Louchard, and Gilles Schaeffer.  \emph{Boltzmann samplers for the random generation
    of combinatorial structures.}  \newblock {\em Combinatorics, Probability {\&}
    Computing}, 13(4-5):577--625, 2004.
\item
  \begin{sloppypar}
    \cite{DBLP:conf/dagstuhl/Tromp06} John Tromp.  \emph{Binary lambda calculus and
      combinatory logic.}  \newblock In Marcus Hutter, Wolfgang Merkle, and Paul
    M.~B. Vit{\'a}nyi, editors, {\em Kolmogorov Complexity and Applications}, volume
    06051 of {\em Dagstuhl Seminar Proceedings}. Internationales Begegnungs- und
    Forschungszentrum fuer Informatik (IBFI), Schloss Dagstuhl, Germany, 2006.
  \end{sloppypar}

\end{enumerate}

The first paper established what is now called de Bruijn indices which play an
important role here in counting $`l$-terms up to $ `a$-conversion.  But Nicolaas de
Bruijn is better known outside the Functional Programming community as one of the
pioneer of modern combinatorics~\cite{bruijn58:_asymp_method_analy}
(see~\cite{knuth00:_selec_paper_analy_algor} fourth cover pages).  We will use
indirectly his combinatorics results here.

The above quotation of Claessen and Hughes is extracted from the introduction
of the second paper~\cite{DBLP:conf/icfp/ClaessenH00} and is completed
by\footnote{See appendix for the whole quotation.}
\begin{it}
  \begin{quotation} We have chosen to put distribution under the human tester's
    control.
  \end{quotation}
\end{it}
However unlike Claessen and Hughes, the philosophy of our paper is that we have
chosen to put the distribution under a strict mathematical control, based on
Boltzmann samplers and probability theory as introduced in the third
paper~\cite{DBLP:journals/cpc/DuchonFLS04}.  For that we use a counting method
introduced by John Tromp in~\cite{DBLP:conf/dagstuhl/Tromp06}.  This amazing paper
deals with algorithmic complexity theory~\cite{261084} and proposes to replace Turing
machines by $`l$-calculus in this theory. Therefore it provides a very simple and
elegant concrete definition of descriptional complexity (Kolmogorov complexity). For
this Tromp proposes to represent $`l$-terms by binary chains and in this framework he
defines a very small self interpreter of size $210$.  His coding of $`l$-terms in
$\{0,1\}^*$ is as follows\footnote{Not exactly since Tromp starts indices at $0$
  like~\cite{LescannePOPL94} and we starts indices at $1$ like de
  Bruijn~\cite{NGDeBruijn108}.}:
\begin{eqnarray*}
  \widehat{n} &=& 1^n0\\
  \widehat{`l M} &=& 00\widehat{M}\\
  \widehat{M\ N} &=& 01\widehat{M}\widehat{N}
\end{eqnarray*}
Representing $`l$-terms by binary chains confers to them a natural notion of size.
This counting is realistic\footnote{In a previous
  submission~\cite{DBLP:journals/jfp/GrygielL13} where we chose to give de Bruijn indices
  size $0$, a referee said: \emph{``If the authors want to use the de Bruijn
  representation, another interesting experiment could be done: rather than to count
  variables as size 0, they should be counted using their \emph{unary}
  representation.  This would penalize deep lexical scoping, which is not a bad idea
  since 'local' terms are much easier to understand and analyze than deep terms''}. }
as it gives variables a weight depending on the distance from their binder. In other
words a variable which is deep in the stack is heavier (larger) than a shallow one .

Moreover our paper relies on the book of Flajolet and
Sedgewick~\cite{flajolet08:_analy_combin} which is the reference on generating
functions.  Our paper is also the application of a recent article containing results
on counting binary $`l$-terms~\cite{DBLP:journals/corr/GrygielL14}.


In this paper we speak about random generation of $`l$-terms.  This applies in
testing and optimizing functional programming
compilers~\cite{Palka:2011:TOC:1982595.1982615,palka12:_testin_compil}, but it is
clear that this applies to other combinatorial objects having bound variables, like
imperative programs with a block structure~\cite{DBLP:conf/pldi/YangCER11}.

\subsubsection*{A survey of Boltzmann approach}
Recall that a Boltzmann sampler  generates uniformly  random objects with a
tolerance in the size of the generated objects.  In other words, the sampler
generates the objects in a cloud around a given size.
\begin{center}
  \includegraphics{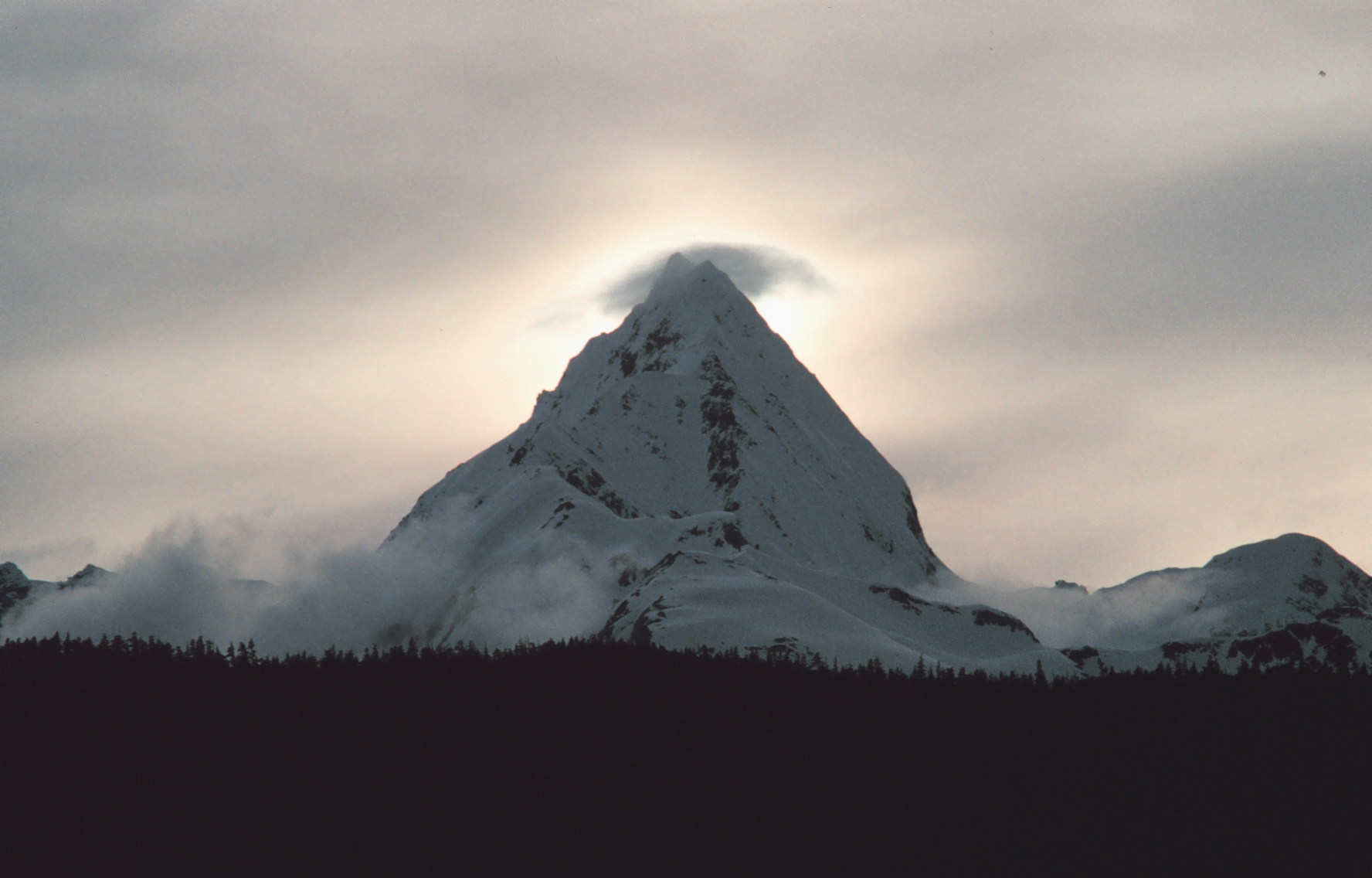}

  {\tiny \it An orographic cloud\footnote{Public Domain Picture, origin: Wikimedia,
      source: \emph{U.S. National Oceanic and Atmospheric Administration}}}
\end{center}

Before a Boltzmann sampler for combinatorial structures is defined, the generating
function associated with these structures is first expressed.  Its coefficients count
the structures by size.  In our case, i.e., trees and $`l$-terms, the expressions of
the generating functions contain a square root of a polynomial, whose smallest root
in module is a real which we call the critical value.  This is the radius of
convergence of the associated series.  A Boltzmann sampler is derived from this
series and based on a value $x$ which allows generating random structures with a mean
and a standard deviation which depend on $x$.  Actually if the user has a specific
size in mind, she provides the desired mean and a calculation returns the $x$ that
yields this mean.  But in our case we assume that we attempt to generate large
structures of size, say at least $100$, perhaps $1000$ or $1000000$ or more. In our
domain, trees and $`l$-terms, the choice of $x$ is not any choice, but $x$ shall be
set to the critical value, for mathematical reasons that are explained
in~\cite{DBLP:journals/cpc/DuchonFLS04}.  This way we get a uniform random sampler
for all big trees or $`l$-terms.

Now let us focus on trees and $\lambda$-terms and assume that the structures are
distributed in kinds. For instance, for $`l$-terms, these kinds are variables,
abstractions and applications.  A random selector among these kinds is built using
probabilities computed using the critical value.  Then the selector is embedded in a
sampler which calls itself recursively.  But if any freedom is left to the sampler,
it may generate huge structures as well as small ones.  To avoid the pitfall of generating
huge objects, the sampler has an upper limit of structures it can
generate. Furthermore after the generation a sieve retains only the structures
that are large enough, yielding the tolerance threshold (the cloud) we mentioned
previously.

\paragraph{Structure of the paper}

The paper is structured as follows.  Since the problem of counting $`l$-terms and
presenting their Boltzmann samplers is not straightforward, we present first in
Section~\ref{sec:trees} the generating functions for trees: binary trees and 1-2
trees. Then we present the generating function for counting $`l$-terms in
Section~\ref{sec:countlbda}.  In Section~\ref{sec:unrank} we present a less efficient
method for generating $`l$-terms not based on Boltzmann samplers. In the following
sections we address Boltzmann samplers. First in Section~\ref{sec:Boltzmann} we
present the general perspective of Boltzmann samplers.  Then in
Section~\ref{sec:BoltzmannTree} we specialize on trees: binary trees and 1-2
trees. Section~\ref{sec:lambda} is devoted to Boltzmann samplers for $`l$-terms.
Until that point the paper addresses only the problem of counting and generating
plain $`l$-terms.  In Section~\ref{sec:closedTerm} we survey the harder problem of
counting and generating closed terms for which we cannot propose a Boltzmann sampler,
since the generating function is not defined by a unique equation.

\section{Counting trees}
\label{sec:trees}

Before presenting how to count $`l$-terms, we will show in this section how to count
trees. We will first show how to count binary trees, then how to count 1-2 trees aka
Motzkin trees.

\subsection{Binary trees}
\label{sec:BinTrees}

A binary tree (the data type \textsf{Tree} in \cite{DBLP:conf/icfp/ClaessenH00})
is either a leaf or a compound tree made of a node and two subtrees. It can be
described as follows:
\[\Bt = \Box \quad + \quad \Bt \times `(!) \times \Bt.\]
We assume that leaves and nodes have size $1$.  General techniques of generating
functions (\cite{flajolet08:_analy_combin} page 738) leads to the following equation
for the generating function \[B(z) = \sum_{n\ge 0} B_n z^n\] whose coefficients count
binary trees of size $n$:
\[B(z) = z + z B(z)^2\] or
\begin{displaymath}
  z B(z)^2 - B(z) + z = 0
\end{displaymath}
which yields:
\[B(z) = \frac{1 - \sqrt{1-4z^2}}{2z}.\] $B(z)$ is not defined for $z`:\real$ larger
that $\frac{1}{2}$. We say that $B(z)$ has a singularity $`r_\Bt$ which is equal to
$\frac{1}{2}$.  We are going to use this fact in our sampler of random binary trees.

\subsection{Motzkin Trees}
\label{sec:MotzTrees}

We are going to consider another family of trees which have analogies with lambda
terms, namely 1-2 trees (\cite{flajolet08:_analy_combin} pages~81). They are also
called \emph{Motzkin trees}. They have two kinds of internal nodes, unary nodes and
binary nodes.  The class of Motzkin trees is described as
\begin{displaymath}
  \M = \Box \quad + \quad `(!) \times \M \quad + \quad \M \times \bullet \times \M.
\end{displaymath}
which means that a Motzkin tree is either a leaf of size $1$ or a tree of size $n+1$
rooted on another tree of size $n$ or a tree of size $n_1 + n_2 +1$ rooted on two
trees of size $n_1$ and~$n_2$.  The generating function 
\begin{displaymath}
  M(z) = \sum_{n=0}^\infty M_n z^n
\end{displaymath}
where $M_n$ counts the number of Motzkin trees of size $n$, is a solution of the
equation:
\begin{displaymath}
  M(z) = z + z M(z) + z M(z)^2
\end{displaymath}
or
\begin{displaymath}
  z M(z)^2 - (1-z)M(z) + z = 0.
\end{displaymath}

Hence
\begin{displaymath}
  M(z) = \frac{1-z - \sqrt{1-2z-3z^2}}{2z}.
\end{displaymath}
It is given by the
smallest root in module of the polynomial $1-2z-3z^2$, namely $`r_{\M}=1/3$ (the
other root $-1$ is larger in module).  
\ifICFP
\begin{figure*}[!t]
  \centering
  \begin{haskell}
    unrankToo :: Int "->" Integer "->" Term\\
    unrankToo n k \\
 \quad\hsalign{| k == (trompoo n) = Index \$  fromIntegral (n - 1)\\
      | k <= (trompoo (n-2)) = Abs (unrankToo (n-2) k) \\
      | \mathbf{otherwise} = unrankApp (n-2) 0 (k - trompoo (n-2))\\
      \textbf{where} unrankApp n j r \hsalign{| r <= tjnj = \hsalign{\textbf{let} (dv,rm) = (r-1) 'divMod' tnj\\
          \mathbf{in} App (unrankToo j (dv+1)) (unrankToo (n-j) (rm+1))}\\
        | \mathbf{otherwise} = unrankApp n (j+1) (r-tjnj)\\
        \textbf{where} \hsalign{tnj = trompoo (n-j)\\
          tjnj = (trompoo j) * tnj}}}
  \end{haskell}
  \caption{The unranking function of $`l$-terms in \textsf{Haskell}}
  \label{fig:unrank}
\end{figure*}
\else
\begin{figure}[!t]
  \centering
  \begin{haskell}
    unrankToo :: Int "->" Integer "->" Term\\
    unrankToo n k \\
 \quad\hsalign{| k == (trompoo n) = Index \$  fromIntegral (n - 1)\\
      | k <= (trompoo (n-2)) = Abs (unrankToo (n-2) k) \\
      | \mathbf{otherwise} = unrankApp (n-2) 0 (k - trompoo (n-2))\\
      \textbf{where} unrankApp n j r \hsalign{| r <= tjnj = \hsalign{\textbf{let} (dv,rm) = (r-1) 'divMod' tnj\\
          \mathbf{in} App \hsalign{(unrankToo j (dv+1))\\
            (unrankToo (n-j) (rm+1))}}\\
        | \mathbf{otherwise} = unrankApp n (j+1) (r-tjnj)\\
        \textbf{where} \hsalign{tnj = trompoo (n-j)\\
          tjnj = (trompoo j) * tnj}}}
  \end{haskell}
  \caption{The unranking function of $`l$-terms in \textsf{Haskell}}
  \label{fig:unrank}
\end{figure}
\fi

\section{Counting lambda terms}
\label{sec:countlbda}

We now consider $`l$-terms up to $`a$-conversion.  To count equivalence classes
modulo $`a$-conversion it is convenient to count canonical representatives of classes
modulo $`a$.  For that the best way is to count terms with de Bruijn indices, since
those terms are unique representatives of classes modulo $`a$.

We work with de Bruijn indices starting at $1$. Sometimes we call those indices
improperly ``variables''.  For instance $`l 1$ is the representative of the
$`a$-conversion equivalence class of the term $`l x.x$ and $`l `l 1~2$ is the
representative of the \mbox{$`a$-conversion} equivalence class of the term $`l x . `l y . y~x$.
We assume that the size of the index $i$ is $i+1$ and the size of an abstraction is
$2$ and the size of an application is $2$ as well.  Therefore we get the equation for
the numbers\footnote{The index $\infty$ is justified by the fact that $S_{\infty,n}$
  is the limit of $S_{m,n}$ when $m$ goes to $\infty$.  See
  Section~\ref{sec:closedTerm}.} of terms of size $n$:
\begin{eqnarray*}
  S_{\infty,0} &=& S_{\infty,1} ~=~ 0,\\
  S_{\infty,n+2} &=& 1 + S_{\infty,n} + \sum_{k=0}^n S_{\infty,k} S_{\infty,n-k}.
\end{eqnarray*}
This means that there is no term of size $0$ or $1$ and for terms of size $n+2$,
there is one term which is an index namely $n+1$, plus $S_{\infty,n}$ terms that are
abstractions plus $\sum_{k=0}^n S_{\infty,k} S_{\infty,n-k}$ terms that are
applications of terms of size $k$ on terms of size $n-k$, for $k$ from $0$ to $n$.
Sequence $(S_{\infty,n})_{n`:\nat}$ can be found in the \emph{On-line Encyclopedia of
  Integer Sequences} with the entry number \textbf{A114851}. Its first $20$ values
are:
\begin{center}
  $0,\ 0,\ 1,\ 1,\ 2,\ 2,\ 4,\ 5,\ 10,\ 14,\ 27,\ 41,$

  $78,\ 126,\ 237,\ 399,\ 745,\ 1292,\ 2404,\ 4259.$
\end{center}
Recall some results of Grygiel-Lescanne~\cite{DBLP:journals/corr/GrygielL14}.  Let $\Soo(z)$
denote the generating function for the sequence $(S_{\infty,n})_{n`:\nat}$, that is
\begin{eqnarray*}
\Soo(z) &=& \sum_{n=0}^{\infty}S_{\infty,n} z^n\\
&=& \sum_{n=0}^{\infty}z^{n+2} + \sum_{n=0}^{\infty} S_{\infty,n} z^{n+2}  \\
&&\quad + \sum_{n=0}^{\infty}\left( \sum_{k=0}^n S_{\infty,k} S_{\infty,n-k}\right)z^{n+2}
\end{eqnarray*}

$\Soo(z)$ fulfills the equation
\[\Soo(z) = \frac{z^2}{1-z} + z^2 \Soo(z) + z^2 \Soo(z)^2.\]
which yields %
\ifICFP the function
\[\frac{(1-z^2)(1-z) - \sqrt{z^6 + 2\,z^5 - 5\,z^4 + 4\,z^3 - z^2 - 2\,z + 1}}{2z^2(1
  - z)}.\]
\else
\[\Soo(z) = \frac{ z^3 - z^2 - z + 1 - \sqrt{z^6 + 2\,z^5 - 5\,z^4 + 4\,z^3 - z^2 -
    2\,z + 1}}{2z^2(1 - z)}.\] %
\fi 
The radius of convergence of this series i.e., the
smallest singularity of the associated analytic function, given as the smallest real
root of the polynomial
\begin{eqnarray*}
  z^6 + 2\,z^5 - 5\,z^4 + 4\,z^3 - z^2 - 2\,z + 1 &=&\\
  (z-1)(z^5 + 3 z^4 - 2 z^3 + 2 z^2 + z - 1).
\end{eqnarray*}
A computation using the mathematical software
\textsf{Sage}~\cite{sage} tells us that this smallest root is approximately
\[{`r_{S_{\infty}}= 0.509308127}.\] Knowledge of the singularity $`r_{S_{\infty}}$
(which we will write just $`r$ in what follows) is crucial for computing quantities
related to the sequence $(S_{\infty,n})_{n`:\nat}$.  Especially it plays a key role
in computing parameters of the Boltzmann sampler.  The following theorem is
in~\cite{DBLP:journals/corr/GrygielL14}.
\begin{theorem}[Asymptotic evaluation of $S_{\infty,n}$]~\\
  The number of all binary \mbox{$`l$-terms} of size $n$ satisfies
  \[ S_{\infty,n} \sim (1/\rho)^n \cdot \frac{C}{n^{3/2}},\] where $1/\rho \doteq
  1.963447954$ and $C \doteq 1.021874073$.
\end{theorem}
It says that whereas binary words are of size $2^n$, binary \mbox{$`l$-terms} are
approximately of size $1.963^n$ for large values of~$n$.  For the skeptic reader, we
have checked numerically this result with good accuracy~\cite{lescanne13}.

\section{Unranking lambda terms}
\label{sec:unrank}

We know how to count the $`l$-terms, which means basically that we know how to order
the terms, assigning a number to each term (ranking).  We may use this fact the other
way around and build a $`l$ term from its rank.  This function is called
\emph{unranking} and is implemented in \textsf{Haskell} as \<unrankToo\> (see
Figure~\ref{fig:unrank}).  The function \<trompoo n\>\footnote{The notation $oo$
  mimics $\infty$.} implements $S_{\infty,n}$.  It works as follows. Given a number
$k$ ($1\le k \le S_{\infty,n}$) we consider which interval it belongs to.  If $k =
S_{\infty,n}$, then $k$ corresponds to a de Bruijn index, namely the de Bruijn index
$n-1$. If $k \le S_{\infty,n-2}$, then $k$ corresponds to an abstraction, namely the
abstraction of the term obtained by unranking $k$ in the interval
$[1..S_{\infty,n-2}]$.  If $S_{\infty,n-2} < k<S_{\infty,n}$, then $k$ corresponds to
an application of two terms which are then computed.

Unranking has at least three applications:
\begin{itemize}
\item counting typable $`l$-terms,
\item generating random $`l$-terms and
\item generating random
  typable $`l$-terms.
\end{itemize}

\subsection{Counting typable $`l$-terms}
\label{sec:cont_typ}

We know very little about the combinatorial properties of simply typable $`l$-terms.
For instance, we do not know any formula for counting them, we have no idea of their
distribution among plain $`l$-terms and we have no information on their ratio among
plain $`l$-terms.  We propose to count how many terms of size $n$ are typable.  We
proceed as follows, we generate all the plain terms of size $n$ and we filter those
that are typable and we count them.  This process is relatively tedious and is
limited by the number of plain terms. Here are the numbers $\nbt{n}$ of typable terms
of size $n$ we computed:
\begin{displaymath}
  \begin{array}{|l|l|}
    \hline
    \mathbf{n} & \nbt{n}\\\hline\hline
    0&0\\\hline
    1&0\\\hline
    2&1\\\hline
    3&1\\\hline
    4&2\\\hline
    5&2\\\hline
    6&3\\\hline
    7&5\\\hline
    8&8\\\hline
    9&13\\\hline
    10&22\\\hline
    11&36\\\hline
    12&58\\\hline
    13&103\\\hline
    14&177\\\hline
    15&307\\\hline
    16&535\\\hline
    17&949\\\hline
    18&1645\\\hline
    19&2936\\\hline
    20&5207\\\hline
  \end{array}
  \qquad\qquad
  \begin{array}{|l|l|}
    \hline
    \mathbf{n} & \nbt{n}\\\hline\hline
    21&9330\\\hline
    22&16613\\\hline
    23&29921\\\hline
    24&53588\\\hline
    25&96808\\\hline
    26&174443\\\hline
    27&316267\\\hline
    28&572092\\\hline
    29&1040596\\\hline
    30&1888505\\\hline
    31&3441755\\\hline
    32&6268500\\\hline
    33&11449522\\\hline
    34&20902152\\\hline
    35&38256759\\\hline
    36&70004696\\\hline
    37&128336318\\\hline
    38&235302612\\\hline
    39&432050796\\\hline
    40&793513690\\\hline
    41 & 1459062947\\\hline
    42& 2683714350\\\hline
  \end{array}
\end{displaymath}
We did not go further than $42$. Notice that
\begin{displaymath}
  S_{\infty,42} = 7\,395\,529\,009 \qquad \textrm{and} \qquad
  S_{\infty,43} = 14\,023\,075\,765.
\end{displaymath}
In words this means that there are more than seven billions of terms of size $42$ and about
$14$ billions of terms of size $43$ which we have to traverse to know how many are
simply typable.  We conjecture~\cite{lescanne13} that like the $S_{\infty,n}$'s, the
$\nbt{n}$'s increase like $1.963447954^n$ and the ratio $\nbt{n} / S_{\infty,n}$ is
polynomial.

\subsection{Generating random $`l$-terms (plain and typable)}
\label{sec:gen}

Thanks to unranking we have a very natural method for generating random
$`l$-terms~\cite{nijenhuis78:_combin} (called by Duchon et
al. \cite{DBLP:journals/cpc/DuchonFLS04} the \emph{recursive method}).  It works as
follows: to generate a random $`l$-terms of size $n$ one generates first a random
number say $k$ in the interval $[1..S_{\infty,n}]$ and then by unranking $k$ one
creates the associated $`l$-term. The size of the term we can generate this way is
limited by the size of the number $S_{\infty,n}$ we can manipulate. 

To generate a random typable $`l$-term of size $n$, one repeats the process of
generating plain terms until one gets a typable $`l$-term.  The process is
inefficient for two reasons: one has to repeat the generation of plain terms with a
lot of waste and one has to handle huge numbers.  But unlike Boltzmann samplers,
unranking generates random $`l$-terms of a precise size. However in applications like
compiler testing, we seldom need to generate terms of a very precise size. An
approximation on the size is enough.

\section{Boltzmann samplers}
\label{sec:Boltzmann}

In this section we present the notion of Boltzmann samplers. For more detail the
reader is invited to look at~\cite{DBLP:journals/cpc/DuchonFLS04}.

A \emph{Boltzmann sampler} returns a random object in a given class $\C$ according to
a given index $x$.  By convention a Boltzmann sampler for $\C$ depending on the index
$x$ is written $`G C(x)$.  In \textsf{Haskell} we will write it \<gammaC gen x\>, because
we need a generator of random numbers to be passed as parameter.

Before specializing the notion of Boltzmann sampler to count specifically lambda
terms of size $n$ assume we consider a generic class $\C$ of combinatorial
structures, where objects of size $n$ are counted by numbers $C_n$.  Assume in
addition that the  class $\C$ admits the generating function $C$, which means that
\begin{displaymath}
  C(x) = \sum_{n=0}^{\infty} C_n x^n.
\end{displaymath}
Uniform probability distribution assigns to each $`g`:\C_n$ the probability:
\[\PP_{\C_n}(`g) = \frac{1}{C_n}\]
We want to generate random objects with some flexibility on the size. In other words
we want the objects to be generated in some cloud around a given size $n$, that is so
that the size $N$ of the objects lies in some interval $(1-`e)n \le N \le
(1+\varepsilon)$ for some factor $\varepsilon>0$ which we call ``tolerance''. Such a
method is called the \emph{approximate-size} uniform random generation.

The \textit{Boltzmann models} assign to any object $`g`:\C$ the following
probability:
\begin{displaymath}
  \PP_{\C_n,x}(`g) = \frac{1}{C(x)}`. x^{|`g|}.
\end{displaymath}

The size of an object in a Boltzmann model is a random variable $N$.  The probability
of drawing an object of size $n$ under the model of index $x$ is
\begin{displaymath}
  \PP_{\C_n}(N=n) = \frac{C_n x^n}{C(x)}.
\end{displaymath}
This is a probability since
\begin{displaymath}
  \sum_{n\ge0} \PP_{\C_n}(N=n) = \frac{1}{C(x)}\sum_{n\ge 0} C_n x^n = 1
\end{displaymath}

The random variable $N$ has \emph{a first moment} and \emph{a second moment}
\cite{DBLP:journals/cpc/DuchonFLS04}:
\begin{displaymath}
  \Me_x(N) = x\frac{C'(x)}{C(x)} \qquad \Me_x(N^2) = \frac{x^2 C''(x) + x C'(x)}{C(x)}.
\end{displaymath}

and a \emph{standard deviation}:
\begin{eqnarray*}
  `s_{\C_n}(x) &=& \sqrt{\Me_x(N^2) - \Me_x(N)^2} \\
  &=& \sqrt{\frac{x^2 C''(x) + x
      C'(x)}{C(x)} - x^2\frac{C'(x)^2}{C(x)^2}}
\end{eqnarray*}
One builds a Boltzmann generator for a class $\C$ according to the recursive
construction of the class~$\C$.

\subsubsection*{Disjoint union}

If the class is built as $\C = \mathscr{A} + \mathscr{B}$, then $C_n = A_n + B_n$ and
$C(z) = A(z) + B(z)$.  The probability of the occurrence of some object is
\begin{displaymath}
  \PP_{\C,x}(`g`:\mathscr{A}) = \frac{A(x)}{C(x)}, \qquad \PP_{\C,x}(`g`:\mathscr{B)} = \frac{B(x)}{C(x)}
\end{displaymath}
Assume we have defined a \textsf{Haskell} data type
\begin{haskell}
  \textbf{data} KindC = IsA | IsB
\end{haskell}
A generator for a Bernoulli variable is given by the \textsf{Haskell} function:
\begin{haskell}
  bern :: Double "->" StdGen "->" (KindC,StdGen)\\
  bern x gen = \hsalign{\textbf{let} (p, g) = randomR (0,1) gen\\
    \mathbf{in} \hsalign{\mathbf{if} p < A(x)/C(x) \mathbf{then} (IsA, g)\\
      \mathbf{else} (IsB,g)}}
\end{haskell}

The class $\C$ is represented by the type \<TypeC\>:
\begin{haskell}
  \textbf{data} TypeC = A TypeA | B TypeB
\end{haskell}
where \<TypeA\> and \<TypeB\> are two types.\footnote{It can be the case that
  \<TypeA\> and \<TypeB\> are \<TypeC\> itself or are defined recursively using
  \<TypeC\>.}  Assume two functions \<gammaA\> and \<gammaB\>.  Then
\begin{normalsize}
  \begin{haskell}
    gammaC :: StdGen "->" Double "->" (C,StdGen)\\
    gammaC gen x = \hsalign{\textbf{let} (class,g) = bern x gen\\
      \mathbf{in} \textbf{case} class \textbf{of}\\
      \hsalign{IsA "->" \hsalign{\textbf{let} (a,g1) = gammaA g x\\
          \mathbf{in} (A a,g1)}}\\
      {IsB "->" \hsalign{\textbf{let} (b,g1) = gammaB g x\\
          \mathbf{in} (B b,g1)}}}
  \end{haskell}
\end{normalsize}
In what follows we will also consider not only a disjoint union of two classes $\C =
\mathscr{A} + \mathscr{B}$, but a disjoint union of three classes like $\C =
\mathscr{A} + \mathscr{B} + \mathscr{D}$.  We call  $randSel$ the function
which replaces $bern$ in this case.

\subsubsection*{Cartesian product}

If the class is built as $\C = \mathscr{A} \times \mathscr{B}$, then the generating
function satisfies $C(z) = A(z) \cdot B(z)$, since
\begin{displaymath}
  C(z) = \sum_{\langle`a,`b\rangle `: \mathscr{A} \times \mathscr{B}} z^{|`a|+|`b|}.
\end{displaymath}
$`g=\langle`a,`b\rangle`:\C$ has the probability:
\begin{displaymath}
  \PP_{\C,x}(`g) = \frac{x^{|`g|}}{C(x)} = \frac{x^{|`a|}}{A(x)}\cdot \frac{x^{|`b|}}{B(x)}.
\end{displaymath}
Assume we have defined a \textsf{Haskell} data type
\begin{haskell}
  \textbf{data} TypeC = Times TypeA TypeB
\end{haskell}
In this case the Boltzmann sampler is
\begin{haskell}
  \hspace*{-10pt} gammaC :: Double "->" StdGen "->" (TypeC,StdGen)\\
  \hspace*{-10pt} gammaC x gen = \hsalign{\textbf{let} \hsalign{(a,g1) = gammaA gen x \\
      (b,g2) = gammaB g1 x}\\
    \mathbf{in} (g2,Times a b)}
\end{haskell}

\subsubsection*{Sequences}
 
$\C$ can be the class of all finite sequences of elements of a class~$\mathscr{A}$. Its
associated type is:
\begin{haskell}
  \textbf{data} TypeC = Nil + Times TypeA TypeC
\end{haskell}
This means that $\C$ is the solution of the equation $$\C = \mathbf{1} + \mathscr{A}
\times \C$$ where $\mathbf{1}$ denotes the empty sequence.  The associated generating
function is solution of $C(z) = 1 + A(z) \cdot C(z)$ which yields:
\begin{displaymath}
  C(z) = \frac{1}{1-A(z)}.
\end{displaymath}
This means that \<Nil\> is selected with probability $1/C(x) = 1-A(x)$ and a non empty
sequence is selected with probability $A(x)$.  
A Boltzmann sampler for sequences is
\begin{haskell}
  \hspace*{-20pt}gammaC  :: Double "->" StdGen "->" (TypeC',StdGen)\\
  \hspace*{-20pt}gammaC x gen = \hsalign{\textbf{let} (p,g) = randomR (0,1) gen\\
    \mathbf{in} \hsalign{\textbf{if} p < A(x)\\
      \textbf{then}  \hsalign{\textbf{let} \hsalign{(a,g1) = gammaA x g\\
          (c,g2) = gammaC x g1}\\
        \mathbf{in} (Times a c,g2)} \\
      \textbf{else} (Nil,g)}}
\end{haskell}

\subsubsection*{The case $x=`r_\C$}

Assume that the generating function we consider is of the form:
\begin{displaymath}
  C(x) = \frac{P_C(x) - \sqrt{Q_C(x)}}{R_C(x)}
\end{displaymath}
where $P_C(x)$, $Q_C(x)$ and $R_C(x)$ are three polynomials and where $`r_\C$ is such that
$Q_C(`r_\C) = 0$ and where $Q_C(x)\neq 0$ and $R_C(x) \neq 0$ for $0\le x\le `r_\C$.  Those properties are
fulfilled for the three generating functions $B(x)$, $M(x)$ and $S_\infty(x)$ we have
seen.  Notice that 
\begin{displaymath}
  C(`r_\C) = \frac{P_C(`r_\C)}{R_C(`r_\C)}
\end{displaymath}
is finite. On another hand
\begin{eqnarray*}
  C'(x) &=& \frac{P'_C(x)}{R_C(x)} - \frac{Q'_C(x)}{2\sqrt{Q_C(x)}R_C(x)} \\
  && \quad - \frac{(P_C(x) - \sqrt{Q_C(x)}) R'_C(x)}{R_C(x)^2}
\end{eqnarray*}
shows that
\begin{displaymath}
  \lim_{x"->"`r_\C} C'(x) = \infty.
\end{displaymath}
Hence 
\begin{displaymath}
  \lim_{x"->"`r_\C} E_x(N) = \lim_{x"->"`r_\C} \frac{x C'(x)}{C(x)} = \infty.
\end{displaymath}
Therefore if we choose $x$ to be  $`r_\C$, the size of the generated structures will
be distributed all over the natural numbers.

\section{Boltzmann samplers for trees}
\label{sec:BoltzmannTree}

\subsection{Boltzmann samplers for Motzkin trees}
\label{sec:Bol-Mot}

Given a number $x$, the \emph{mean value} formula (or \emph{first moment})
$\Me_{x,\M}(N) = x M'(x)/M(x)$ allows us by solving the equation $x M'(x)/M(x) = n$
to tell which value of $x$ provides a sampler with $n$ as mean value for the size.
For instance, $x M'(x)/M(x) = 100$ returns $x_{100} = 0.33330833286456574$ it says
that the sampler \emph{gammaM gen $x_{100}$} (see below) will generate a random term
with a mean value $100$.  Similarly since $x_{600} = 0.3333326388880542$,
\emph{gammaM gen $x_{600}$} will generate a random term with mean value $600$. For a
small value, say $10$ we get $x_{1}= 0.3308286281723805$.  The standard deviation
$`s_{\M_n}(x)$ can be instantiated to $x_{10}$ or $x_{100}$ or to~$x_{600}$.
Actually
\begin{displaymath}
  `s_{\M_n}(x_{10}) \approx 25 \quad `s_{\M_n}(x_{100}) \approx 816 \quad  `s_{\M_n}(x_{600}) \approx 1200.
\end{displaymath}
Added to the fact that $x_{100}$ and $x_{600}$ are very close to the $`r_{\M} = 1/3$
(which is called the \emph{critical value}) it appears that a Boltzmann sampler
for large values of $n$ is obtained by parameterizing $gammaM$ with $`r_{\M}$, namely
$1/3$.  See \cite{DBLP:journals/cpc/DuchonFLS04} Section 7.2 for a mathematical
justification of this choice.

Since there are three components in the addition, we generalize the Bernoulli
generator and then we specialize it to the case of Motzkin trees, producing a
function we call \<randSelM\>.  Let us call \[\frac{M_0(z)}{M(z)}=\frac{z}{M(z)}\]
the first component of $(z + z M(z) + z M(z)^2)/M(z)$, \[\frac{M_1(z)}{M(z)}=z\] the
second component of $(z + z M(z) + z M(z)^2)/M(z)$ and \[\frac{M_2(z)}{M(z)} = z
M(z)\] the third component of $(z + z M(z) + z M(z)^2)/M(z)$.  We define the
following \textsf{Haskell} functions:
\begin{haskell}
  m z = (1-z - sqrt(1-2*z-3*z*z))/(2*z) \\
  p1 x = x / m x \\
  p2 x = p1 x + x
\end{haskell}
One may notice that \<p2 x + x * m x == 1\> for every value of $x$, which means that
when one adds all the probabilities of the events one gets $1$.

\begin{haskell}
  \hspace*{-20pt}\textbf{data} KindMotzkin = IsLeaf | IsUnary | IsBinary \\
  ~\\
  \hspace*{-20pt}randSelM :: StdGen "->"  Double "->" (KindMotzkin,StdGen)\\
  \hspace*{-20pt}randSelM gen x = \hsalign{\textbf{let} (p,g) = randomR (0,1) gen\\
    \mathbf{in} \hsalign{\textbf{if} p < p1 x \textbf{then} (IsLeaf,g)\\
      \textbf{else}  \mathbf{if} p < p2 x  \textbf{then} (IsUnary,g)\\
      \textbf{else} (IsBinary,g)}}
\end{haskell}
Then we define the data type
\begin{haskell}
  \textbf{data} MTree = MNil | MU MTree | MB MTree MTree
\end{haskell}
for Motzkin trees and we get the Boltzmann sampler for Motzkin trees:
\begin{haskell}
  \hspace*{-20pt}gammaM :: StdGen "->" Double "->" (MTree,StdGen) \\
  \hspace*{-20pt}gammaM gen x = \hsalign{\textbf{let} (kind,g) = randSelM gen x\\
    \mathbf{in} \textbf{case} kind \textbf{of}\\
    IsLeaf "->" (MNil, g)\\
    IsUnary "->" \hsalign{\textbf{let} (t1,g1) = gammaM g x\\
      \mathbf{in} (MU t1,g1)}\\
    IsBinary "->" \hsalign{\textbf{let} \hsalign{(t1,g1) = gammaM g x\\
        (t2,g2) = gammaM g1 x}\\
      \mathbf{in} (MB t1 t2, g2)}}
\end{haskell}
An interesting feature should be noticed in this program.  Indeed the function
\<gammaM gen x\> calls \<gammaM g x\> and should not terminate since we can prove in
no way that the first arguments namely \<gen\> versus \<g\> decrease. Thus the
termination is only probabilistic and depends on the choice of $x$.  The
implementation of \<randSelM\> makes \<gammaM gen x\> to terminate with
probability $1$, thanks to the whole theory of Boltzmann samplers. For instance if
the generator is buggy and takes its values only in the interval $(\frac{1}{3},1]$
the program will not terminate.

\ifICFP
\begin{figure*}[!t]
  \centering
  \begin{haskell}
    ceiledGammaM  :: StdGen "->" (EMTree,Int,StdGen)\\
    ceiledGammaM gen = \hsalign{\textbf{let} (kind,g0) = selOneThird gen\\
      \mathbf{in} \textbf{case} kind \mathbf{of}\\
      IsLeaf "->" (OK MNil, 1, g0)\\
      IsUnary "->" \hsalign{\textbf{let} (et1,n1,g1) = ceiledGammaM g0\\
        \mathbf{in} \hsalign{\textbf{case} et1 \mathbf{of}\\
          OK t1 "->" \hsalign{\textbf{if} n1 <= upLimit\\
            \textbf{then} (OK (MU t1), n1+1, g1)\\
            \textbf{else} (Error,upLimit,g1)}\\
          Error "->" (Error,upLimit,g1)}}\\
      IsBinary "->" \hsalign{\textbf{let} \hsalign{(et1,n1,g1) = ceiledGammaM g0\\
          (et2,n2,g2) = ceiledGammaM g1}\\
        \mathbf{in} \hsalign{\textbf{case} et1 \mathbf{of}\\
          OK t1 "->" \hsalign{\textbf{case} et2 \mathbf{of}\\
            OK t2 "->" \hsalign{\textbf{if} n1+n2+1 <=upLimit \\
              \textbf{then} (OK (MB t1 t2), n1+n2+2, g2)\\
              \textbf{else} (Error,upLimit,g2)}\\
            Error "->" (Error,upLimit,g2)}\\
          Error "->" (Error,upLimit,g2)}}}
  \end{haskell}
  \caption{The function $ceiledGammaM$}
  \label{fig:ceiledMotkin}
\end{figure*}
\else
\begin{figure}[!t]
  \centering
  \begin{haskell}
    ceiledGammaM  :: StdGen "->" (EMTree,Int,StdGen)\\
    ceiledGammaM gen = \\
\hspace*{10pt}\hsalign{\textbf{let} (kind,g0) = selOneThird gen\\
      \mathbf{in} \textbf{case} kind \mathbf{of}\\
      IsLeaf "->" (OK MNil, 1, g0)\\
      IsUnary "->" \hsalign{\textbf{let} (et1,n1,g1) = ceiledGammaM g0\\
        \mathbf{in} \hsalign{\textbf{case} et1 \mathbf{of}\\
          OK t1 "->" \hsalign{\textbf{if} n1 <= upLimit\\
            \textbf{then} (OK (MU t1), n1+1, g1)\\
            \textbf{else} (Error,upLimit,g1)}\\
          Error "->" (Error,upLimit,g1)}}\\
      IsBinary "->" \hsalign{\textbf{let} \hsalign{(et1,n1,g1) = ceiledGammaM g0\\
          (et2,n2,g2) = ceiledGammaM g1}\\
        \mathbf{in} \hsalign{\textbf{case} et1 \mathbf{of}\\
          OK t1 "->" \hsalign{\textbf{case} et2 \mathbf{of}\\
            OK t2 "->" \hsalign{\textbf{if} n1+n2+1 <=upLimit \\
              \textbf{then} (OK (MB t1 t2), n1+n2+2, g2)\\
              \textbf{else} (Error,upLimit,g2)}\\
            Error "->" (Error,upLimit,g2)}\\
          Error "->" (Error,upLimit,g2)}}}
  \end{haskell}
  \caption{The function $ceiledGammaM$}
  \label{fig:ceiledMotkin}
\end{figure}
\fi
\subsubsection*{Samplers for large Motzkin trees}
Note that $M(`r_\M) = 1$. For large Motzkin trees the right choice is to take $x =
`r_{\M} = 1/3$. This gives the probability
\begin{itemize}
\item for drawing a leaf $\frac{`r_\M}{M(`r_\M)} = \frac{1}{3}$,
\item for drawing a tree rooted by a unary node $`r_\M = \frac{1}{3}$,
\item for drawing a tree rooted by a binary node ${`r_\M M(`r_\M) = \frac{1}{3}}$.
\end{itemize}
Actually since choosing $`r_\M$ generates large trees,  it is advisable to
limit the size of the tree to be generated during the process.  Therefore, we
consider what Duchon et al. call \emph{ceiled random generation}.  We define a
constant \<upLimit\> which sets a size upper limit that the sampler should respect
during the generation.  We then define a data type:
\begin{haskell}
  data EMTree = OK MTree | Error
\end{haskell}
and a selection function \<selOneThird\> which selects a kind among \<IsLeaf\>,
\<IsUnary\> and \<IsBinary\> with probability $1/3$, $1/3$, $1/3$.  The function
\<ceiledGammaM\> is defined in Figure~\ref{fig:ceiledMotkin}.

If we want to generate a term between say $a$ and $b$ we generate terms with an
\<upLimit\> set to $b$ until we get a term larger than $a$.  Duchon et al. have shown
that the complexity of this method is linear. In other words, the generation of a
tree with a size between $a$ and $b$ takes a time $0(b)$.

\subsection{Boltzmann samplers for binary trees}
The binary trees are implemented in \textsf{Haskell} by
\begin{haskell}
  data BTree = BNil | BNode BTree BTree
\end{haskell}
If we compute for $B(z)$ the means associated to some values of $n$, a we did above
for $M(z)$, we get:
\begin{eqnarray*}
  x_{10} &=& 0.4960783708241233 \\  
  x_{100} &=& 0.4999739685017913 \\ 
  x_{600} &=& 0.49999930090217926
\end{eqnarray*}
for which we obtain the standard deviations:
\begin{displaymath}
  `s_{\Bt_n}(x_{10}) \approx  22 \quad `s_{\Bt_n}(x_{100}) \approx  970 \quad  `s_{\Bt_n}(x_{600}) \approx 14623.
\end{displaymath}
There are two kinds of binary trees:
\begin{itemize}
\item \emph{leafs} (represented by \<BNil\>) which correspond to
  \begin{displaymath}
    \frac{B_0(z)}{B(z)} = \frac{z}{B(z)}
\end{displaymath}
which is the first component of $(z + z B(z)^2)/B(z)$,
  and
\item \emph{nodes} (represented by \<BNode t1 t2\>) which correspond to
  \begin{displaymath}
\frac{B_1(z) }{B(z)} = z B(z)
\end{displaymath}
which is the
  second component of $(z + z B(z)^2)/B(z)$.
\end{itemize}

Let us define
\begin{haskell}
  b z = (1 - sqrt(1 - 4*z*z))/ (2*z)\\
 q x = x / b x
\end{haskell}
We introduce
\begin{haskell}
  data KindBTree = IsBNil | IsBNode
\end{haskell}
and
\begin{haskell}
\hspace*{-20pt}randSelB  :: StdGen "->"  Double "->" (KindBTree,StdGen)\\
\hspace*{-20pt}randSelB gen x = \hsalign{\textbf{let} (p,g) = randomR (0,1) gen\\
           \textbf{in} \hsalign{\textbf{if} p < q x \\
           \textbf{then} (IsBNil,g)\\
              \textbf{else} (IsBNode,g)}}
\end{haskell}
We can then introduce the sampler:
\begin{haskell}
\hspace*{-25pt} gammaB :: StdGen "->" Double "->" (BTree,StdGen)\\
\hspace*{-25pt} gammaB gen x = 
\qquad \hsalign{\textbf{let} (kind,g) = randSelB gen x\\
               \textbf{in} \hsalign{\textbf{case} kind \textbf{of}\\
                  IsBNil "->" (BNil, g)\\
                  IsBNode "->" \hsalign{\textbf{let} \hsalign{(t1,g1) = gammaB g x\\
                                 (t2,g2) = gammaB g1 x\\
                             \hsalign{\textbf{in} (BNode t1 t2, g2)}}}}}
\end{haskell}

Once again we see that the best value for $x$ to generate random large binary trees
is  $`r_{\Bt} = \frac{1}{2}$. Thus \<gammaB gen 0.5\> generates large
binary trees.  For a similar approach of the same problem with in particular a
discussion of \emph{ceiled random generation} see~\cite{yorgey13:_blog}.  For other
methods for generating random binary trees, see~\cite{KnuthVol4_4}, pages~18-19
and~\cite{DBLP:journals/ita/Remy85}.

\section{Boltzmann samplers for lambda terms}
\label{sec:lambda}

Like for Motzkin trees we consider the equation of the generating function:
\[\Soo(z) = \frac{z^2}{1-z} + z^2 \Soo(z) + z^2 \Soo(z)^2.\]
with three components: the first corresponds to de Bruijn indices, the second to
abstractions, the third to applications.  Like for Motzkin trees we have to build a
random selector among three probabilities.  First we describe a data type:
\begin{haskell}
  \hspace*{-25pt} \textbf{data} KindTerm = IsVariable | IsAbstraction | IsApplication
\end{haskell}
a function \<soo\> %
\ifICFP given in Figure~\ref{fig:soo}
\begin{figure*}[t]
  \centering
  \begin{haskell}
 soo z = (z^3 - z^2 - z + 1 - sqrt(z^6 + 2*(z^5) - 5*(z^4) + 4*(z^3) - z^2 - 2*z + 1))/(2*z*z*(1 - z))
  \end{haskell}
  \caption{The \textsf{Haskell} function \emph{soo} implementing $\Soo(z)$}
  \label{fig:soo}
\end{figure*}
\else
\begin{haskell}
soo z = (z^3 - z^2 - z + 1 - sq)/(2*z*z*(1 - z))\\
\hspace*{10pt}\textbf{where} sq = sqrt(z^6 + 2*(z^5) - 5*(z^4) + 4*(z^3) - z^2 - 2*z + 1)
\end{haskell}
\fi %
and two functions:
\begin{haskell}
  p1 x = x*x / (1-x) / soo x\\
  p2 x = p1 x + x^2
\end{haskell}
Using \textsf{Sage} we computed the values:
\begin{eqnarray*}
  x_{100} &=& 0.5092252666102192\\
  x_{500} &=& 0.5093048407797965\\
  x_{600} &=& 0.5093058457062517\\
  x_{1000} &=& 0.5093073063214039
\end{eqnarray*}
which are the choices of $x$ yielding mean values $100$, $500$, $600$ and $1000$ respectively.

\subsubsection*{General samplers of $`l$-terms}
The values of the probabilities for a given $x$ are
\begin{itemize}
\item $p_v(x) = \frac{x^2}{(1-x) S_\infty(x)}$ for variables,
\item $p_{abs}(x) = x^2$ for abstractions,
\item $p_{app}(x) = x^2 S_\infty (x)$ for applications.
\end{itemize}
For selecting among \<IsVariable\>, \<IsAbstraction\> and \<IsApplication\> we use
the following selector written in \textsf{Haskell}
\begin{haskell}
  \hspace*{-10pt}randSel :: StdGen "->" Double "->" (KindTerm,StdGen)\\
  \hspace*{-10pt}randSel gen x = \hsalign{\textbf{let} (p,g) = randomR (0,1) gen\\
    \mathbf{in} \hsalign{\textbf{if} p < pv x \textbf{then} (IsVariable,g)\\
      \textbf{else} \hsalign{\textbf{if} p < pv x + pabs x\\
        \textbf{then} (IsAbstraction,g)\\
        \textbf{else} (IsApplication,g)}}}
\end{haskell}

\subsubsection*{Samplers for large $`l$-terms}

Like for trees,  the best choice of $x$ for generating large
$`l$-terms is $\rho$ which we call \<rho\> in \textsf{Haskell}:
\begin{haskell}
  rho :: Double\\
  rho = 0.5093081270242373.
\end{haskell}
whose square is
\begin{haskell}
  rhosquare = rho * rho
\end{haskell}
which yields $`r^2 = 0.25939476825293667$.  Notice that since $`r$ is a root of the
polynomial below the square root, $S_\infty(`r) = \frac{1-`r^2}{2`r^2}$.  The values
of the probabilities for selecting among variables, abstractions and applications
are:
\begin{itemize}
\item $p_v(`r) = \frac{2`r^4}{(1-`r)(1-`r^2)}$ for variables,
\item $p_{abs}(`r) = `r^2$ for abstractions,
\item $p_{app}(`r) = \frac{1-`r^2}{2}$ for applications.
\end{itemize}
Let us simplify $\frac{2`r^4}{(1-`r)(1-`r^2)}$ into $\frac{1-`r^2}{2}$ by computing
the difference:
\begin{eqnarray*}
  \frac{2`r^4}{(1-`r)(1-`r^2)} - \frac{1-`r^2}{2} &=& \frac{4`r^4 -
    (1-`r^2)^2(1-`r)}{2(1-`r)(1-`r^2)}\\
  &=& \frac{`r^5 + 3`r^4 -2`r^3 + 2`r^2 + `r -1}{2(1-`r)(1-`r^2)}\\
  &=& 0.
\end{eqnarray*}
Therefore we get the result announced in the introduction namely that
\begin{itemize}
\item $p_v(`r) = \frac{1-`r^2}{2} \approx 0.3703026$ for variables,
\item $p_{abs}(`r) = `r^2 \approx 0.25939476$ for abstractions,
\item $p_{app}(`r) = \frac{1-`r^2}{2} \approx 0.3703026$ for applications.
\end{itemize}
We build the random selector:
\begin{haskell}
  \hspace*{-10pt}pvrho = (1 - rhosquare)\\
  \hspace*{-10pt}p2rho = pvrho + rhosquare\\
  ~\\
  \hspace*{-10pt}randSelRho :: StdGen "->" (KindTerm,StdGen)\\
  \hspace*{-10pt}randSelRho gen = \hsalign{\textbf{let} (p,g) = randomR (0,1) gen\\
    \mathbf{in} \hsalign{\textbf{if} p < pvrho \textbf{then} (IsVariable,g)\\
      \textbf{else} \hsalign{\textbf{if} p < p2rho \\
        \textbf{then} (IsAbstraction,g)\\
        \textbf{else} (IsApplication,g)}}}
\end{haskell}
We generate a Boltzmann sampler for indices (or variables).  A de Bruijn index is a
natural number i.e.,  a sequence of~$1$'s.  Hence we take the Boltzmann sampler for
sequences with $A(z)=z$ and $A(`r) = `r$.
\begin{haskell}
  gammaV :: StdGen "->" (Integer, StdGen)\\
  gammaV gen = \hsalign{\textbf{let} (a,g) = randomR (0,1) gen\\
    \mathbf{in} \hsalign{\textbf{if} a < rho\\
      \textbf{then} \hsalign{\textbf{let} (n,g1) = gammaV g\\
        \mathbf{in} (n+1,g1) \\
        \textbf{else} (1,g)}}}
\end{haskell}
We want to generate terms that are below a certain \<upLimit\>.  Thus when the limit
is passed we generate an error.
For that we create a
type with error.
\begin{haskell}
  data ETerm = OK Term | Error
\end{haskell}
This method recalls the \textbf{sized} generation
of~\cite{DBLP:conf/icfp/ClaessenH00}. %
\ifICFP A Boltzmann sampler for large $`l$-terms
is given in Figure~\ref{fig:gammaSoo}.
\begin{figure*}
  \centering
  \begin{haskell}
    ceiledGammaSoo :: StdGen"->" (ETerm,Int,StdGen)\\
    ceiledGammaSoo gen = \hsalign{\textbf{let} (kind,g0) = randSelRho gen\\
      \textbf{in} \mathbf{case} kind \textbf{of} \\
      \hsalign{IsVariable "->"  \hsalign{\textbf{let} (n,g1) = gammaV g0\\
          \textbf{in}  \hsalign{\textbf{if} fromIntegral n <= upLimit \\
            \textbf{then} (OK (Index n), fromIntegral (n+1),g1)\\
            \textbf{else} (Error,upLimit, g1)}}\\
        IsAbstraction "->"  \hsalign{\textbf{let} (et1,n1,g1) = ceiledGammaSoo g0\\
          \textbf{in}  \hsalign{\textbf{case} et1 \textbf{of}\\
            OK t1 "->"  \hsalign{\textbf{if} n1+2 <= upLimit \\
              \textbf{then} (OK (Abs t1),n1+2,g1) \\
              \textbf{else} (Error,upLimit, g1)}\\
            Error  "->" (Error,upLimit, g1)}}    \\
        IsApplication "->"  \hsalign{\textbf{let}  \hsalign{(et1,n1,g1) = ceiledGammaSoo g0\\
            (et2,n2,g2) = ceiledGammaSoo g1}\\
          \textbf{in}  \hsalign{\textbf{case} et1 \textbf{of}\\
            OK t1 "->"  \hsalign{\textbf{case} et2 \textbf{of} \\
              OK t2 "->"  \hsalign{\textbf{if} n1+n2+2 <= upLimit \\
                \textbf{then} (OK (App t1 t2),n1+n2+2,g2) \\
                \textbf{else} (Error,upLimit, g2)}\\
              Error "->" (Error,upLimit, g2)}\\
            Error "->" (Error,upLimit, g2)}}}}
  \end{haskell}
  \caption{Boltzmann sampler for large $`l$-terms}
  \label{fig:gammaSoo}
\end{figure*}
\else
A Boltzmann sampler for large $`l$-terms which is given:
\begin{haskell}
\hspace*{-20pt}  ceiledGammaSoo :: StdGen"->" (ETerm,Int,StdGen)\\
  ceiledGammaSoo gen = \\
\hspace*{10pt}\hsalign{\textbf{let} (kind,g0) = randSelRho gen\\
    \textbf{in} \mathbf{case} kind of \\
    \hsalign{IsVariable "->"  \hsalign{\textbf{let} (n,g1) = gammaV g0\\
        \textbf{in}  \hsalign{\textbf{if} fromIntegral n <= upLimit \\
          \textbf{then} (OK (Index n), fromIntegral (n+1),g1)\\
          \textbf{else} (Error,upLimit, g1)}}\\
      IsAbstraction "->"  \hsalign{\textbf{let} (et1,n1,g1) = ceiledGammaSoo g0\\
        \textbf{in}  \hsalign{case et1 of\\
          OK t1 "->"  \hsalign{\textbf{if} n1+2 <= upLimit \\
            \textbf{then} (OK (Abs t1),n1+2,g1) \\
            \textbf{else} (Error,upLimit, g1)}\\
          Error  "->" (Error,upLimit, g1)}}    \\
      IsApplication "->"  \hsalign{\textbf{let}  \hsalign{(et1,n1,g1) = ceiledGammaSoo g0\\
          (et2,n2,g2) = ceiledGammaSoo g1}\\
        \textbf{in}  \hsalign{case et1 of\\
          OK t1 "->"  \hsalign{case et2 of \\
            OK t2 "->"  \hsalign{\textbf{if} n1+n2+2 <= upLimit \\
              \textbf{then} (OK (App t1 t2),n1+n2+2,g2) \\
              \textbf{else} (Error,upLimit, g2)}\\
            Error "->" (Error,upLimit, g2)}\\
          Error "->" (Error,upLimit, g2)}}}}
\end{haskell}
\fi

To generate a large plain $`l$-term, terms are filtered until a term that is large
enough is generated.  Recall that the method is linear in time complexity.  Thus the
generation of a term of size $100,000$ takes a few seconds, the
generation of a term of size one million takes three minutes and the generation of a
term of size five millions takes five minutes on a laptop.

To generate large typable $`l$-terms we generate $`l$-terms and check their
typability. Currently we are able to generate random typable $`l$-terms of size
$500$.

\section{Ranking and unranking closed lambda terms}
\label{sec:closedTerm}

Boltzmann samplers work only when one knows the generating function explicitly.  Thus
if one wants to generate typable closed terms, one has two solutions.  First one
generates a random typable term using the Boltzmann sampler described above and if
the produced term is not closed, one closes it by adding the adequate number of
abstractions.  Second one uses an unranking method similar to this described in
Section~\ref{sec:gen}.  For that let us recall how one counts terms with at most $m$
free indices~\cite{DBLP:journals/corr/GrygielL14}. The specific case of the closed terms is
the case when the number of free indices is at most~$0$.
\begin{eqnarray*}
  S_{m,0} &=& S_{m,1} ~=~ 0,\label{eq:Smn}\\
  S_{m,n+2} &=& [m \ge n+1] + S_{m+1,n} + \sum_{k=0}^n S_{m,k} S_{m,n-k}.\label{eq:Smn2}
\end{eqnarray*}
Given a predicate $P$, $[P(\vec{x})]$ denotes the Iverson symbol, i.e., $[P(\vec{x})]
= 1$ if $P(\vec{x})$ and $[P(\vec{x})] = 0$ if $\neg P(\vec{x})$.
Figure~\ref{fig:unrankT-prog} gives a program for unranking terms with at most $m$ indices,
\ifICFP
\begin{figure*}[!th]
  \begin{center}
    \begin{normalsize}
      \begin{haskell}
        unrankT :: Int "->" Int "->" Integer "->" Term\\
        unrankT m n k \\
\hspace*{10pt}\hsalign{| m >= n - 1 \&\& k == (tromp m n) = Index \$ fromIntegral (n - 1) \hscom{terms $1^{n-1}0$}\\
          | k <= (tromp  (m+1) (n-2)) = Abs (unrankT (m+1) (n-2) k) \hscom{terms $00M$}\\
          | \mathbf{otherwise} = unrankApp (n-2) 0 (k - tromp  (m+1) (n-2)) \hscom{terms $01MN$}\\
               \hspace*{10pt}\textbf{where} unrankApp n j h \hsalign{| h <=  tmjtmnj  = \hsalign{\textbf{let}(dv,rm) = (h-1) 'divMod' tmnj\\
              \mathbf{in} App (unrankT m j (dv+1)) (unrankT  m (n-j) (rm+1))}\\
            | \mathbf{otherwise} = unrankApp n (j + 1) (h -tmjtmnj) \\
            \hspace*{10pt}\textbf{where} \hsalign{tmnj = tromp m (n-j)\\
              tmjtmnj = (tromp m j) * tmnj }}}
      \end{haskell}
    \end{normalsize}
  \end{center}
  \caption{\emph{unrankT} function in \textsf{Haskell}.}
  \label{fig:unrankT-prog}
\end{figure*}
\else
\begin{figure*}[!th]
  \begin{center}
    \begin{normalsize}
      \begin{haskell}
      \hspace*{-20pt}
        unrankT :: Int "->" Int "->" Integer "->" Term\\
        unrankT m n k \\
\hspace*{10pt}\hsalign{| m >= n - 1 \&\& k == (tromp m n) = Index \$ fromIntegral (n - 1) \hscom{terms $1^{n-1}0$}\\
          | k <= (tromp  (m+1) (n-2)) = Abs (unrankT (m+1) (n-2) k) \hscom{terms $00M$}\\
          | \mathbf{otherwise} = unrankApp (n-2) 0 (k - tromp  (m+1) (n-2)) \hscom{terms $01MN$}\\
               \hspace*{10pt}\textbf{where} unrankApp n j h \\
\hspace*{20pt}\hsalign{| h <=  tmjtmnj  = \hsalign{\textbf{let}(dv,rm) = (h-1) 'divMod' tmnj\\
              \mathbf{in} App (unrankT m j (dv+1)) (unrankT  m (n-j) (rm+1))}\\
            | \mathbf{otherwise} = unrankApp n (j + 1) (h -tmjtmnj) \\
            \hspace*{10pt}\textbf{where} \hsalign{tmnj = tromp m (n-j)\\
              tmjtmnj = (tromp m j) * tmnj }}}
      \end{haskell}
    \end{normalsize}
  \end{center}
  \caption{\emph{unrankT} function in \textsf{Haskell}.}
  \label{fig:unrankT-prog}
\end{figure*}
\fi
where \<tromp m n\> is the \textsf{Haskell} function representing $S_{m,n}$.
$S_{0,n}$ counts closed terms and is computed by \<tromp 0 n\>.

\section{Related works}
\label{sec:works}

In the introduction we cited papers that are clearly connected to this work.  In a
recent work, Bacher et al.~\cite{DBLP:journals/corr/BacherBJ14} propose an improved
random generation of binary trees and Motzkin trees, based on R\'{e}my algorithm~\cite{DBLP:journals/ita/Remy85} (or
algorithm R as Knuth calls it~\cite{KnuthVol4_4}).  Instead of growing the trees from the root, they
propose like R\'{e}my to grow the trees from inside by an operation called grafting.
It is not clear how this can be generalized to $`l$-terms as this requires ``to find a
combinatorial interpretation for the holonomic equations [which] is not [...]
always possible, and even for simple combinatorial objects this is not elementary''
(Conclusion of \cite{DBLP:journals/corr/BacherBJ14} page~16).  

We would like also to mention papers on counting
\mbox{$`l$-terms} \cite{DBLP:journals/tcs/Lescanne13,DBLP:journals/jfp/GrygielL13} and
evaluating their combinatorial proprieties
namely \cite{bodini11:_lambd_bound_unary_heigh,DBLP:journals/corr/abs-0903-5505,DBLP:journals/combinatorics/BodiniGGJ13,DBLP:journals/tcs/BodiniGJ13}.
Another related paper is~\cite{DBLP:conf/haskell/DuregardJW12} which proposes
\textsf{Haskell} programs for enumerating structures.

\section{Acknowledgments}

We would like to thank Katarzyna Grygiel, Bruno Salvy and John Tromp,
for fruitful discussions.  We are also indebted to Neil Sloane and the developers of
the \emph{On-line Encyclopedia of Integer Sequences} which allowed us to know about
sequence \textbf{A114851} which was the starting point of this research.

\section{Conclusion}

Boltzmann samplers are central tools for the uniform generation of random
structures, like trees or $`l$-terms.  Two directions are now open for applications:
first to integrate the programs proposed here in actual testers and optimizers and
second to extend Boltzmann samplers to other kinds of programs, for instance programs
with block structures. From the theoretical point of view, more should be known about
generating functions for \emph{closed $`l$-terms} or \emph{$`l$-terms with fixed bound on the
number of variables}.  Boltzmann samplers should be designed for such terms which
requires to extend the theory. Concerning combinatorial properties of \emph{simply typable
$`l$-terms} many question are left open and seem to be hard. Besides since we are interested
in generating typable terms, it could be worth to build random uniform samplers
 delivering directly typable terms, for instance based on a hashing table built offline.

\appendix
Here is the whole quotation of Claessen and Hughes~\cite{DBLP:conf/icfp/ClaessenH00}:

\begin{quotation}
  \begin{it}
    We have chosen to put distribution under the human tester's control, by defining
    a test data generation language (also embedded in Haskell), and a way to observe
    the distribution of test cases. By programming a suitable generator, the tester
    can not only control the distribution of test cases, but also ensure that they
    satisfy arbitrarily complex invariants.  
  \end{it}
\end{quotation}

\end{document}